\documentstyle[12pt]{article}

\def\beq{\begin{equation}}
\def\eeq{\end{equation}}
\def\bea{\begin{eqnarray}}
\def\eea{\end{eqnarray}}

\def\ba{\begin{array}}
\def\ea{\end{array}}

\def\,{\"{U}}
\def\6{\.{I}}
\begin{document}
\title{Exact solution of Schr\"{o}dinger equation for modified Kratzer's
molecular potential with the position-dependent mass}
\author{\vspace{1cm}
{Ramazan Sever $^1$, Cevdet Tezcan $^2$ }
         \\
{\small \sl  $^1$ Middle East Technical University,Department of
Physics, 06531 Ankara, Turkey }
\\{\small \sl $^2$Faculty of Engineering, Ba\c{s}kent University, Ba\~{g}l{\i}ca
Campus, Ankara, Turkey  }}
\date{\today}
\maketitle
\begin{abstract}

Exact solutions of Schr\"{o}dinger equation are obtained for the
modified Kratzer and the corrected Morse potentials with the
position-dependent effective mass. The bound state energy
eigenvalues and the corresponding eigenfunctions are calculated for
any angular momentum for target potentials. Various forms of point
canonical transformations are applied.\\
PACS numbers: 03.65.-w; 03.65.Ge; 12.39.Fd\\
Keywords: Morse potential, Kratzer potential, Position-dependent
mass, Point canonical transformation, Effective mass Schr\"{o}dinger
equation.
\end{abstract}

\newpage

\section{Introduction}
Solutions of Schr\"{o}dinger equation for a given potential with any
angular momentum have much attention in chemical physics systems.
Energy eigenvalues and the corresponding eigenfunctions provide a
complete information about the diatomic molecules.

Morse and Kratzer potentials [1,2] are one of the well-known
diatomic potentials. The method used in the Schr\"{o}dinger equation
for vibration-rotation states are mostly based on the wave function
expansion and exact solution for a single state with some
restrictions on the coupling constants [3-7]. On the other hand
solutions of the position-dependent effective-mass Schr\"{o}dinger
equation are very interesting chemical potential problem.

They have also found important applications in the fields of
material science and condensed matter physics such as
semiconductors[8], quantum well and quantum dots[9], $^{3}H$,
clusters[10], quantum liquids[11], graded alloys and semiconductor
heterostructures]12,13].

Recently, number of exact solutions on these topics
increased[14-31]. Various methods are used in the calculations. The
point canonical transformations (PCT) is one of these methods
providing exact solutions of energy eigenvalues and corresponding
eigenfunctions [24-27]. It is also used for solving the
Schr\"{o}dinger equation with position-dependent effective mass for
some potentials [8-13].

In the present work, we solve two different potentials with the
three mass distributions. The point canonical transformation is
taken in the more general form introducing a free parameter. This
general form of the transformation will provide us a set of
solutions for different values of free parameter. In this work, the
exact solution of Schr\"{o}dinger equation is obtained or the
modified Kratzer type of molecular potential [31] and the corrected
Morse potential [32].

The contents of the paper is as follows. In section 2, we present
briefly the solution of the Schr\"{o}dinger by using point canonical
transformation. In section 3, we introduce some applications for the
specific mass distributions. Results are discussed in section 4.

\section{Method}
To introduce the PCT, we start from a time independent
Schr\"{o}dinger equation for a potential $V(y)$

\begin{equation}
\left(-\frac{1}{2}\frac{d^2}{dx^2}+V(y)\right)\phi(y)=E\phi(y)
\end{equation}
where the atomic unit $\hbar=1$ and the constant mass $M=1$ are
taken. Defining a transformation $y\rightarrow x$ for a mapping
$y=f(x)$, the wave function can be rewritten as

\begin{equation}
\phi(y)=m(x)\psi(x).
\end{equation}
The transformed Schr\"{o}dinger equation takes

\begin{eqnarray}
&&\left\{-\frac{1}{2}\frac{d^{2}}{dx^{2}}-\left(\frac{m^{\prime}}{m}-
\frac{f^{\prime\prime}}{2f^{\prime}}\right)\frac{d}{dx}-\frac{1}{2}\left[\frac{m^{\prime\prime}}{m^{\prime}}
+(\alpha-1)\left(\frac{m^{\prime}}{m}\right)^{2}-\left(\frac{m^{\prime}}{m}\right)
\frac{f^{\prime\prime}}{f^{\prime}}\right]\right.\nonumber\\[0.3cm]
&+&\left.\left(f^{\prime}\right)^{2}~V(f(x))\right\}~\psi(x)=(f^{\prime})^{2}E~\psi(x),
\end{eqnarray}
where the prime denotes differentiation with respect to $x$. On the
other hand the one dimensional Schr\"{o}dinger equation with
position dependent mass can be written as

\begin{equation}
-\frac{1}{2}\frac{d}{dx}\left[\frac{1}{M(x)}\frac{d\psi(x)}{dx}\right]+\tilde{V}(x)\psi(x)=\tilde{E}\psi(x),
\end{equation}
where $M(x)=m_{0}~m(x)$, and the dimensionless mass distribution
$m(x)$ is real function. For simplicity, we take $m_{0}=1$. Thus,
Eq. (4) takes the form

\begin{equation}
\left(-\frac{1}{2}\frac{d^2}{dx^2}+\frac{m^\prime}{2m}\frac{d}{dx}+m\tilde{V}(x)\right)\psi(x)=m\tilde{E}\psi(x).
\end{equation}
Comparing Eqs. (3) and (5), we get the following identities

\begin{equation}
\frac{f^{\prime\prime}}{2f^\prime}-\frac{m^\prime}{m}=\frac{m^\prime}{2m}
\end{equation}
and

\begin{equation}
\tilde{V}(x)-\tilde{E}=\frac{{f^{\prime}}^{2}}{m}\left[V(f(x))-E\right]-
\frac{1}{2m}\left[\frac{m^{\prime\prime}}{m}-
\left(\frac{m^\prime}{m}\right)\left(\frac{f^{\prime\prime}}{f^\prime}\right)\right]
\end{equation}
From Eq. (6), one gets

\begin{equation}
f^\prime=m^{1/2}
\end{equation}
Substituting $f^{\prime}$ into Eq. (7), the new potential can be
obtained as

\begin{equation}
\tilde{V}(x)=V(f(x))-\frac{1}{8m}\left[\frac{m^{\prime\prime}}{m}
-\frac{7}{4}\left(\frac{m^{\prime}}{m}\right)^{2}\right].
\end{equation}
Therefore, the energy eigenvalues and corresponding wave functions
for the potential $V(y)$ as $E_n$ and $\phi_{n}(y)$ become

\begin{eqnarray}
\tilde{E}_n&=&E_n
\end{eqnarray}
and

\begin{eqnarray}
\psi_n(x)&=&\frac{1}{m(x)}\phi_n(y)
\end{eqnarray}

\section{Applications}
We solve Schr\"{o}dinger equation exactly for two potentials the
rotationally corrected Morse potential[30] and the modified Kratzer
molecular potential[31]. We consider three kinds of the position
dependent mass distributions. Two of them are used before[19], and
the third one is the exponentially decreasing mass distribution with
a free parameter q.

\subsection{Modified Kratzer Potential}
\begin{equation}
V(r)=D_{e}\left(\frac{y-y_{e}}{y}\right)^{2},\label{eq1}
\end{equation}
where $D_{e}$ is the dissociation energy and $y_{e}$ is the
equilibrium internuclear separation. Energy spectrum and the wave
functions are

\begin{equation}
E_{n\ell}(n)=D_{e}-\frac{\hbar^{2}}{2\mu}\left[\left(\frac{4\mu
D_{e}y_{e}}{\hbar^{2}}\right)^2\left(1+2n+\sqrt{1+4\left(\frac{2\mu
D_{e}y_{e}^{2}}{\hbar^{2}}+\ell(\ell+1)\right)}\right)^{-2}\right],
\end{equation}

\begin{equation}
R_{n\ell}=A_{n\ell}(2i\varepsilon
y)^{-\frac{1}{2}(1-\eta)}e^{-i\varepsilon
y}L_{n}^{\sqrt{1+4\gamma}}(2i\varepsilon y)
\end{equation}
where

\begin{equation}
\eta=\sqrt{1+4\gamma},
\end{equation}

\begin{equation}
\gamma=\frac{2\mu\left(D_{e}y_{e}^{2}+\frac{\ell(\ell+1)\hbar^{2}}{2\mu}\right)}{\hbar^{2}},
\end{equation}

\begin{equation}
A_{n\ell}=\left(\frac{8\mu
D_{e}y_{e}}{\hbar^{2}(2n+\eta+1)}\right)^{3/2}\left[\frac{n!}{(2n+\eta+1)(n+\eta)!}\right]^{1/2},
\end{equation}

\begin{equation}
\epsilon=\frac{i \beta}{(2n+\eta+1)},
\end{equation}
and

\begin{equation}
\beta=-\frac{4\mu D_{e}y_{e}}{\hbar^{2}},
\end{equation}

\subsubsection{Asymptotically vanishing mass distribution $m(x)=\frac{a^2}{q+x^{2}}$}

\begin{equation}
y=f(x)=\int m(x)^{1/2}dx=a\ell n(x+\sqrt{q+x^{2}}),
\end{equation}
The target potential is

\begin{equation}
\tilde{V}(x)=D_{e}\left[\frac{a\ell n(x+\sqrt{q+x^{2}})-y_{e}}{a\ell
n(x+\sqrt{q+x^{2}})}\right]^{2}-\frac{1}{8a^{2}}\frac{2q+x^{2}}{q+x^{2}},
\end{equation}
Energy eigenvalues and the normalized radial wave function for the
target potential $\tilde{V}(x)$ are

\begin{equation}
\tilde{E}_{n}=E_{n\ell}(n),
\end{equation}

\begin{equation}
R_{n\ell}(x)=A_{n\ell}(2i\varepsilon \ell
n(x+\sqrt{q+x^{2}}))^{-\frac{1}{2}(1-\eta)}e^{-i\varepsilon \ell
n(x+\sqrt{q+x^{2}})}L_{n}^{\sqrt{1+4\gamma}}(2 i a\varepsilon \ell
n(x+\sqrt{q+x^{2}}))
\end{equation}

\begin{equation}
A_{n\ell}^{2}=\frac{4a
n!(1+n+\varepsilon_{1})^{2}(2\sqrt{\varepsilon_{3}})^{2\varepsilon_{1}}}{(1+n+2\varepsilon_{1})!}.
\end{equation}

\subsubsection{Asymptotically vanishing mass distribution $m(x)=\frac{a^2}{(b+x^{2})^{2}}$}

\begin{equation}
y=f(x)=\int
m(x)^{1/2}dx=\frac{a}{\sqrt{b}}\tan^{-1}\frac{x}{\sqrt{b}},
\end{equation}
The target potential

\begin{equation}
\tilde{V}(x)=D_{e}\left(\frac{\frac{a}{\sqrt{b}}\tan^{-1}\frac{x}{\sqrt{b}}
-y_{e}}{\frac{a}{\sqrt{b}}\tan^{-1}\frac{x}{\sqrt{b}}}\right)^{2}-\frac{1}{2a^{2}}(b+2x^{2}),
\end{equation}
and the corresponding energy spectrum and the wave function are

\begin{equation}
\tilde{E}_{n}=E_{n\ell}(n),
\end{equation}

\begin{equation}
R_{n\ell}(x)=A_{n\ell}\left(2i\varepsilon
\frac{a}{\sqrt{b}}\tan^{-1}\frac{x}{\sqrt{b}}\right)^{-\frac{1}{2}(1-\eta)}e^{-i\varepsilon
\frac{a}{\sqrt{b}}\tan^{-1}\frac{x}{\sqrt{b}}}L_{n}^{\sqrt{1+4\gamma}}(2i\varepsilon
\frac{a}{\sqrt{b}}\tan^{-1}\frac{x}{\sqrt{b}}).
\end{equation}

\subsubsection{Exponentially vanishing mass distribution $m(x)=e^{-qx}$}

\begin{equation}
y=f(x)=\int m(x)^{1/2}dx=-\frac{2}{q}e^{-\frac{q}{2}x},
\end{equation}

\noindent the target potential

\begin{equation}
\tilde{V}(x)=D_{e}\left(1+\frac{1}{2}q
r_{e}e^{\frac{q}{2}x}\right)^{2}+\frac{9}{128}q^{4}e^{-qx},
\end{equation}
and the corresponding energy spectrum and the wave function are

\begin{equation}
\tilde{E}_{n}=E_{n\ell}(n),
\end{equation}

\begin{equation}
R_{n\ell}(x)=A_{n\ell}\left(-\frac{4}{\alpha}i\varepsilon
e^{-\frac{\alpha}{2}x}\right)^{-\frac{1}{2}(1-\eta)}e^{\frac{2}{\alpha}i\varepsilon
e^{-\frac{\alpha}{2}x}}L_{n}^{\sqrt{1+4\gamma}}(-\frac{4}{\alpha}i\varepsilon
e^{-\frac{\alpha}{2}x}).
\end{equation}

\subsection{Rotationally corrected Morse Potential}

\begin{equation}
V(y)=D(e^{-2\alpha y}-2e^{-\alpha y})+\gamma(D_{0}+D_{1}e^{-\alpha
y}+D_{2}e^{-2\alpha y}),
\end{equation}

\begin{equation}
\alpha=a r_{0},
\end{equation}

\begin{equation}
\gamma=\frac{\hbar^{2}\ell(\ell+1)}{2\mu r_{0}^{2}},
\end{equation}
$r_{0}$ is the equilibrium intermolecular distance, $a$ is a
parameter controlling the width of the potential wall. $D$ is the
dissociation energy and

\begin{equation}
D_{0}=1-\frac{3}{\alpha}+\frac{3}{\alpha^{2}},
\end{equation}

\begin{equation}
D_{1}=\frac{4}{\alpha}-\frac{6}{\alpha^{2}},
\end{equation}

\begin{equation}
D_{2}=-\frac{1}{\alpha}+\frac{3}{\alpha^{2}},
\end{equation}
Energy spectrum and the radial wave function are

\begin{equation}
E_{n\ell}=\frac{\hbar^{2}\ell(\ell+1)}{2\mu
r_{0}^{2}}\left(1-\frac{3}{a
r_{0}}+\frac{3}{a^{2}r_{0}^{2}}\right)-\frac{\hbar^{2}a^{2}}{2\mu}\left[\frac{\varepsilon_{2}}{2\sqrt{\varepsilon_{3}}}-(n+\frac{1}{2})\right]^{2},
\end{equation}
where

\begin{equation}
\frac{\varepsilon_{2}}{2\sqrt{\varepsilon_{3}}}=\frac{1}{a^{2}\sqrt{\varepsilon_{3}}}\left[\frac{2\mu
D}{\hbar^{2}}-\frac{\ell(\ell+1)}{r_{0}^{2}}\left(\frac{2}{a
r_{0}}-\frac{3}{a^{2}r_{0}^{2}}\right)\right],
\end{equation}

\begin{equation}
R_{n\ell}(y)=A_{n\ell}e^{-\alpha\varepsilon_{1}y}
e^{-\sqrt{\varepsilon_{3}}e^{-\alpha y}}
L_{n}^{1+2\varepsilon_{1}}(2\sqrt{3}e^{-\alpha y}),
\end{equation}

\begin{equation}
-\varepsilon_{1}^{2}=\frac{2\mu r_{0}^{2}(E_{n\ell}-\gamma
D_{0})}{\hbar^{2}\alpha^{2}},
\end{equation}

\begin{equation}
-\varepsilon_{2}=\frac{2\mu r_{0}^{2}(2D-\gamma
D_{1})}{\hbar^{2}\alpha^{2}},
\end{equation}

\begin{equation}
-\varepsilon_{3}=\frac{2\mu r_{0}^{2}(D+\gamma
D_{2})}{\hbar^{2}\alpha^{2}},
\end{equation}

\subsubsection{Asymptotically vanishing mass distribution $m(x)=\frac{a^2}{q+x^{2}}$}
$y$ is given in 3.1.1. The target potential

\begin{equation}
\tilde{V}(x)=\frac{D+\gamma D_{1}}{(x+\sqrt{q+x^{2}})^{2\alpha
a}}+\frac{\gamma D_{1}-2D_{2}}{(x+\sqrt{q+x^{2}})^{\alpha a}}+\gamma
D_{0}
\end{equation}
Energy spectrum and the wave function are

\begin{equation}
\tilde{E}_{n}=E_{n\ell}(n),
\end{equation}

\begin{equation}
R_{n\ell}(x)=\frac{A_{n\ell}}{(x+\sqrt{q+x^{2}})^{\alpha\varepsilon_{1}
a}}e^{-\frac{\sqrt{\varepsilon_{3}}}{(x+\sqrt{q+x^{2}})^{\alpha
a}}}L_{n}^{1+2\varepsilon_{1}}\left(\frac{2\sqrt{\varepsilon_{3}}}{(x+\sqrt{q+x^{2}})^{\alpha
a}}\right).
\end{equation}

\subsubsection{Asymptotically vanishing mass distribution $m(x)=\frac{a^2}{(b+x^{2})^{2}}$}
The target potential

\begin{equation}
\tilde{V}(x)=(D+\gamma D_{1})e^{-\frac{2\alpha
a}{\sqrt{b}}\tan^{-1}\frac{x}{\sqrt{b}}}+(\gamma
D_{1}-2D)e^{-\frac{\alpha
a}{\sqrt{b}}\tan^{-1}\frac{x}{\sqrt{b}}}+\gamma
D_{0}-\frac{q+2x^{2}}{2a^{2}}
\end{equation}
Energy spectrum and the wave function are

\begin{equation}
\tilde{E}_{n}=E_{n\ell}(n),
\end{equation}

\begin{equation}
R_{n\ell}(x)=A_{n\ell}e^{-\varepsilon_{1}P{x}}
e^{-\sqrt{\varepsilon_{3}}e^{-
P{x}}}L_{n}^{1+2\varepsilon_{1}}\left(2\sqrt{\varepsilon_{3}}~e^{-P{x}}\right).
\end{equation}

\subsubsection{Exponentially vanishing mass distribution $m(x)=e^{-qx}$}
The target potential

\begin{equation}
\tilde{V}(x)=D\left(e^{\frac{4\alpha}{q}e^{-\frac{qx}{2}}}
-2e^{\frac{2\alpha}{q}e^{-\frac{qx}{2}}}\right)+\gamma\left(D_{0}
+D_{1}e^{\frac{2\alpha}{q}e^{-\frac{qx}{2}}}+D_{2}e^{\frac{4\alpha}{q}e^{-\frac{qx}{2}}}\right),
\end{equation}
Energy spectrum and the wave function

\begin{equation}
\tilde{E}_{n}=E_{n\ell}(n),
\end{equation}

\begin{equation}
R_{n\ell}(x)=A_{n\ell} [T(x)]^{\epsilon_{1}} e^{-\sqrt{\epsilon_{3}}
T(x)}L^{1+2\epsilon_{1}}_{n}\left(2\sqrt{3}T(x)\right)
\end{equation}
where $T(x)=e^{Q(x)}$ and $Q(x)=\frac{2\alpha}{q}e^{-\frac{q}{2}x}$.

\section{Conclusions}
We have applied the PCT in a general form by introducing a free
parameter to solve the Schr\"{o}dinger equation for the corrected
Morse and modified Kratzer potentials with spatially dependent mass.
In the computations, we have used three position dependent mass
distributions. Energy eigenvalues and corresponding wave funtions
for target potentials are written in the compact form.

\section{Acknowledgements}

This research was partially supported by the Scientific and
Technological Research Council of Turkey.

\end{document}